\documentclass{article}

\usepackage{arxiv}

\usepackage[utf8]{inputenc} 
\usepackage[T1]{fontenc}    
\usepackage{hyperref}       
\usepackage{url}            
\usepackage{booktabs}       
\usepackage{amsfonts}       
\usepackage{nicefrac}       
\usepackage{microtype}      
\usepackage{lipsum}		
\usepackage{graphicx}
\usepackage[numbers]{natbib}
\usepackage{doi}
\usepackage{multirow}
\usepackage{adjustbox}
\usepackage{rotating}
\usepackage{multirow}
\usepackage{caption}
\usepackage{lscape}

\title{Quantifying Indirect Gender Discrimination on Collaborative Platforms}

\author{ \href{https://orcid.org/0000-0001-6326-0617}{\includegraphics[scale=0.06]{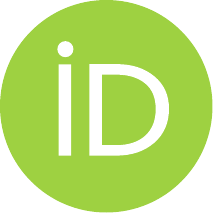}\hspace{1mm}Orsolya V\'as\'arhelyi}\thanks{Use footnote for providing further
		information about author (webpage, alternative
		address)---\emph{not} for acknowledging funding agencies.} \\
	Center for Collective Learning\\
	Corvinus Institute for Advanced Studies, \\ 
        Institute of Data Analytics and Information Systems \\
        Corvinus University of Budapest\\
	Budapest, Hungary 1093 \\
	\texttt{orsolya.vasarhelyi@uni-corvinus.hu} \\
	\And
	\href{https://orcid.org/0000-0002-9469-8774}{\includegraphics[scale=0.06]{orcid.pdf}\hspace{1mm}Bal\'azs Vedres} \\
	Department of Network and Data Science\\
	Central European University\\
	Vienna, Austria, 1100 \\
        Oxford Internet Institute \\
        University of Oxford \\
        Oxford, United Kingdom, OX1 3JS \\
	\texttt{vedresb@ceu.edu } \\
}

\renewcommand{\shorttitle}{Quantifying Indirect Gender Discrimination on Collaborative Platforms}

\hypersetup{
pdftitle={A template for the arxiv style},
pdfsubject={q-bio.NC, q-bio.QM},
pdfauthor={David S.~Hippocampus, Elias D.~Striatum},
pdfkeywords={First keyword, Second keyword, More},
}

\begin{document}
\maketitle

\begin{abstract}
Digital collaborative platforms have become crucial venues of career advancement and individual success in many creative fields, from engineering to the arts. Indirect gender discrimination is a key component to gendered disadvantage on platforms. Such platforms carried the promise of opening avenues of advancement to previously discriminated groups, such as women, as platforms lack managerial gatekeepers with conventional prejudice. We analyzed the extent of indirect gender discriminatory on two diverse platforms, GitHub and Behance, focused on software development and fine arts and design. We found that the main cause of women's disadvantage in attention, success, and survival is largely due to indirect discrimination that varies between 60-90\% of total female disadvantage. Men and women are penalized if they follow highly female-like behavior, while categorical gender's impact varies by outcome and field. As platforms employ algorithmic tools and AI systems to manage users' activity, visibility and recommend new projects to collaborate, stereotypes rooted in behavior can have long-lasting consequences.
\end{abstract}


\section{Introduction}

Platform organizations offer digital affordances to connect producers and consumers, and this organisational form had seen a rapid uptake over the past decade. Today, the world’s most valuable businesses (Apple, Microsoft, Amazon, Google, or Facebook) are platforms \citep{stark2020algorithmic, de2017impact}, and a recent report estimated that global digital platforms in 2022 had 371 billion average monthly users \citep{dinarstandard}. Platformization does not appear to slow down, as the annualized growth rate of digital trades (20\%) is faster than that of physical products (6\%) \citep{stojkoski2023growth}. This acceleration has resulted in an entire new ecosystem \citep{magaudda2020platform}, which has changed the way we communicate \citep{magaudda2020platform}, shop \citep{xue2020literature}, travel \citep{schor2016debating, pelzer2019institutional, cansoy2020homines}, define success \citep{magaudda2020platform}, work \citep{schor2020dependence, stark2020algorithmic} and collaborate \citep{mergel2015open,gray2016crowd, moldon2021}.

Digital collaborative platforms have become crucial tools for independent creative workers, providing opportunities to develop skills, connect with other like-minded people and potential collaborators, and help capture the attention of potential users and buyers \citep{grabher2020disruption, churchill2019gender}. GitHub, for example, is a platform where programmers can find collaborators, start projects, be noticed by potential team members and employers, and thus succeed \citep{Bonaccorsi2003, Dabbish2012, borges2018s}. Other platforms, such as Behance, play a similar role for graphic artists, allowing them to build open access portfolios of their visual work, find peers and collaborators, and ultimately catch the attention of buyers and clients \citep{halstead2015finding, scolere2019brand}. 

The emergence of such portfolio careers \citep{neff2005entrepreneurial} 
carried the promise of extending opportunities to previously disadvantaged groups, among them women \citep{van2020inclusivity,ford2016paradise,huws2019hassle}. However, this reputation-based new economy translates social capital and risk taking behavior into value which appears to benefit men more than women \citep{poutanen2017new, bogliacino2019quantity, silberzahn2014pay, cook2018gender}. Platforms often lack features that take gender inequalities into account, and, furthermore, the deepest aspects of their culture, technology, design, and algorithmic management tend to perpetuate gender discrimination \citep{Terrell2017GenderMen,wachs2017, wachs2018and, may2019gender, vedres2019gendered, brooke2021trouble}. 

Several recent publications set out to chart the gender gap in the digital economy with regards to participation and success \citep{vasilescu2015gender, wagner2015s, wagner2016women, horvat2017gender, ford2017someone}. A key critical point raised about these works is that gender itself is already encoded in the creation of technology (including the algorithms that govern digital platforms), and thus platforms would not be able to mitigate gender inequalities \citep{grau2021gender, vasarhelyi2022computing}. Gender as a social concept assumes that men and women follow their category-specific scripts: norms and behaviors that reinforce societal expectations \citep{west1987doing}. This {\it gendered behavior} is learned through socialization and individuals are penalized if they deviate from it \citep{lorber2018paradoxes}. Femininity and masculinity are often in a hierarchical relationship; in the field of technology it is the superiority of men and masculine culture that is is considered natural and is thus rewarded.

Gender discrimination can be either {\it direct} or {\it indirect}. Direct discrimination occurs when decisions and processes are based on an individual's categorical gender identity, resulting in disadvantages by category. Gender discrimination can also be indirect, based on nonsensitive attributes -- like activity patterns or specializations --, that are closely linked to gender. There is evidence of direct \citep{Terrell2017GenderMen} and indirect discrimination against women on technical platforms \citep{vedres2019gendered,may2019gender,brooke2024programmed}, despite the fact that there is no difference in the quality of code produced by men and women \citep{Terrell2017GenderMen,brooke2024programmed}. 

Previous studies found \citep{vedres2019gendered,may2019gender,brooke2024programmed} that one's gender can be predicted fairly accurately based on their collaboration patterns, specialization, and style of code they produce. A prior study of GitHub \citep{vedres2019gendered} operationalized "femaleness" as the extent of feminine behavior by this resulting probability. Their results showed that men and women are both penalized if they follow highly female-like behavior, indicating the presence of indirect gender discrimination.

Studies of various digital platforms (with varying proportions of female participants) have found that users' gender can be predicted based on their online activity. In a recent study users' gender was predicted on Pinterest, where the ratio of women is equal to men, based on the content they share \citep{mittal2014pinned}. A study of the digital music platform The Echo Nest \citep{wang2019gender}, found that 25\% of the solo artists are women, and the authors managed to predict the artist's gender based on the musical features of their songs.  

If users' activity on digital platforms differ markedly by gender, does this also result in indirect gender discrimination? Wachs and coauthors \citep{wachs2018and} found that gender differences in the popularity and visibility of projects can be explained by the "gendereness" of skills and the structure of the follower network in the Dribble designer community. Their study indicates that even within a community where every fourth active user is a woman (compared to every 10th on GitHub) indirect gender discrimination is a problem. 

This article contributes to quantifying indirect gender discrimination in the platform economy \citep{wachs2017,vedres2019gendered, may2019gender} in three major ways. First, we provide a multi-platform analysis, as we replicate prior work \citep{vedres2019gendered} on Behance, a platform that is on the opposite end of the creative spectrum in terms of content to GitHub. Behance is a platform for visual arts and graphic design that emphasizes visual creativity. This platform features individual creatives who display portfolios of their photographs, paintings, logos, or other applied visual work. Despite significant differences in the content, all key aspects of programmers' activity on GitHub could be transferred to the context of graphic artists on Behance -- likely due to isomorphism in platform design. This enables us to make more general statements about gender inequality on platforms, that a single-platform study would not warrant.  

Second, we re-operationalize gendered behavior to closely capture indirect discrimination. \citep{wachs2017,vedres2019gendered, wang2019gender} showed that gendered tie formation is related to success and can improve models to predict one's gender. However, gender differences in the number of men and women with whom one collaborates can be the result of direct discrimination. Women might collaborate less with men because men simply do not find them "worthy" to work with \citep{windsor2015femininities}. To avoid mixing the effects of direct and indirect discrimination, we have removed the gendered aspects of collaborations from the prediction model that captures "femaleness".

Finally, third, we add an outcome measure, attention, to augment two measures used before, popularity, and survival. Attention precedes popularity and survival as an initial form of success, in the sense that platform users first need to be noticed before they can succeed along any other dimensions \citep{wachs2017}. Studies of gender inequalities point to inequalities in being noticed in the workplace as a key dimension of women's disadvantage. Women in highly masculine professions face a paradoxical visibility problem: they attract considerable attention as women, but this does not translate into their acceptance as experts \citep{Fernando2018, wang2018competence}. In other words, we expect less direct discrimination in attention (as women often attract attention, especially in fields where they are underrepresented), while it is an open question whether there is indirect discrimination by gender-typical behavior in gaining attention. 

In sum, we found that both men and women are penalized if they follow highly female-like behavior: On collaborative platforms indirect discrimination is the main source of gender disparities. This pattern holds for attention, success, and survival, and it is true for both GitHub and Behance. Our findings should be especially alarming, as self-learning algorithms are fast becoming widespread on platforms, managing users' activity and visibility. Indirect gender discrimination presents a grave risk of a deeply rooted, invisible, and stubborn inequality, as it can be baked into the algorithms and culture of online collaborations, greatly magnifying already existing gender inequalities.

\section{Methods}

\subsection{Data}

GithHb (www.github.com) is by far the most popular collaborative platform for software projects. It offers online hosting and version control services, that allows developers to contribute to software projects from around the world. According to Octoverse, the annual statistical report of GitHub, the platform had more than 100 million user accounts in 2023, regardless of their activity status \citep{Octoverse2023}. Since GitHub provides affordances beyond the recording of contributions and the management of the source code -- such as traditional social media functionalities (e.g., following) --, it became the subject of several studies, aiming to understand collaborative activity online \citep{gousios2017mining, zoller2020topology}, team success, diversity \citep{vasilescu2015gender, ortu2017diverse}, and gender inequality in technology \citep{imtiaz2019investigating}.

In this article, we use a dataset obtained from githubarchive.org, containing individual careers between 2009-02-19 and 2016-10-21. This dataset contains the following information for each individual: the creation of a repository, push to a repository, opening, closing, and merging pull requests; accompanied by users’ information using the GitHub API. (user names, email addresses, number of followers, number of public repositories and date they joined GitHub; see Table \ref{tab:vars}). 

Behance (www.behance.net) is a collaborative platform for creative professionals, where they can feature a portfolio of their work, collect, and organize the works of others for inspiration, and become hired as freelance workers. Similarly to GitHub, Behance allows users to create relationships via social network features (following, commenting, and appreciating) and share their work in a wide variety of domains such as photography, graphic design, and user experience (UX) research. Behance is a considerably smaller platform than GitHub, with about 50 million users (according to Behance.net). 

The original data source of our study is a randomized sample of the Behance database obtained by Kim, 2017 \citep{kim2017creative} accompanied by additional data points using the official Behance API (https://www.behance.net/dev). The data contained information on the gender of 37,777 users, specialized topics, number of followers, number of users followed, number of appreciations (likes), number of comments, project views,  and stylistic information of the projects, and the total number of projects. We used the official Behance API, to collect more detailed user information about the date of registration, the activity status, and the users' names. This allowed us to evaluate the results of the applied gender inferring method.

In online collaborative platforms users often create accounts without aiming to maintain an active presence. (Users might open an account out of curiosity or to access some affordances outside of creative work, like digital storage.) Therefore, in both databases, we filter users by the level of activity, retaining only those users with at least 10 traces of activity within their careers. Furthermore, we removed users with names containing substrings that classified them as potential artificial agents (e.g.: "bot", "test", "daemon", "svn2github", "gitter-badger"). In the case of Behance, we removed all users whose accounts we could not connect with the API anymore and did not have a display name, which is an indication of being a company. The resulting database contains 1,634,373 GitHub users and 30,186 Behance users. 

\subsection*{Gender Inferring}

Since none of the data sources lists users' gender, we infer gender from publicly available name information listed by users: first and last names, email addresses, nicknames. Inferring users' perceived gender from public name data based on large-scale gender-name dictionaries has been widely used in computational social science \citep{vasarhelyi2022computing, shugars2024categorizing}. However, it is important to note that these methods can also introduce bias into our results. They perform considerably better with Western names, compared to Asian ones \citep{karimi2016, lockhart2023name} and usually produce binary gender categories \citep{shugars2024categorizing}. These are important limitations that must be taken into account when discussing results \citep{shugars2024categorizing}, however, we believe that gender-inferring algorithms that attempt to mimic how humans decide about the gender identity of users are valid methods for the purpose of our study. Our study focuses on how {\it perceived gender} impacts user outcomes on collaborative online platforms, and since the public tends to be biased and prefers to categorize people into gender groups \citep{shaw2014internet}, our automated method can serve as a suitable proxy for understanding gender inequalities. 

In the case of GitHub, we infer first names from display names, usernames, and e-mail addresses using the methods developed by \citep{vedres2019gendered}. Behance data was published with inferred gender, using a commercial service called Gender API (https://gender-api.com/). Table \ref{tab:gender_inf} shows the resulting database by data source and gender. Gender recognition on GitHub yields 11.87\% females and 88.13\% males out of all users with names, while on Behance the resulting database contained 29.45\% females and 70.55\% males. After filtering for users with at least 10 traces of activity on both platforms, on GitHub the ratio of female users decreases to 5.49\%, while on Behance the ratio of active female users decreases to 28.39\%. 

\begin{table}[ht]
\centering
\begin{tabular}{lll}
\toprule
                        & GitHub    & Behance \\
\midrule
N in population         & 7,798,509 & 37,777  \\
Female                  & 194,000   & 11,124  \\
Male                    & 1,441,130 & 26,653  \\
Unknown                 & 6,163,379 &    -    \\
\midrule
N After Filtering      & 1,634,373 & 30,186  \\
Female                  & 56,731    & 8,569   \\
Male                    & 977,389   & 21,617  \\
Unknown                 & 600,253   &   -     \\
Sample size (by gender) & 10,000    & 6,000   \\
\bottomrule
\end{tabular}
\caption{Data cleaning and gender inferring results. After filtering for users with at least 10 activity points, in GitHub the ratio of female users is 5.49\%, and on Behance the ratio of active female users is 28.39\% out of those users whose gender could be inferred.}
\label{tab:gender_inf}
\end{table}

\begin{figure*}[ht]
\centering
\includegraphics[width=.9\linewidth]{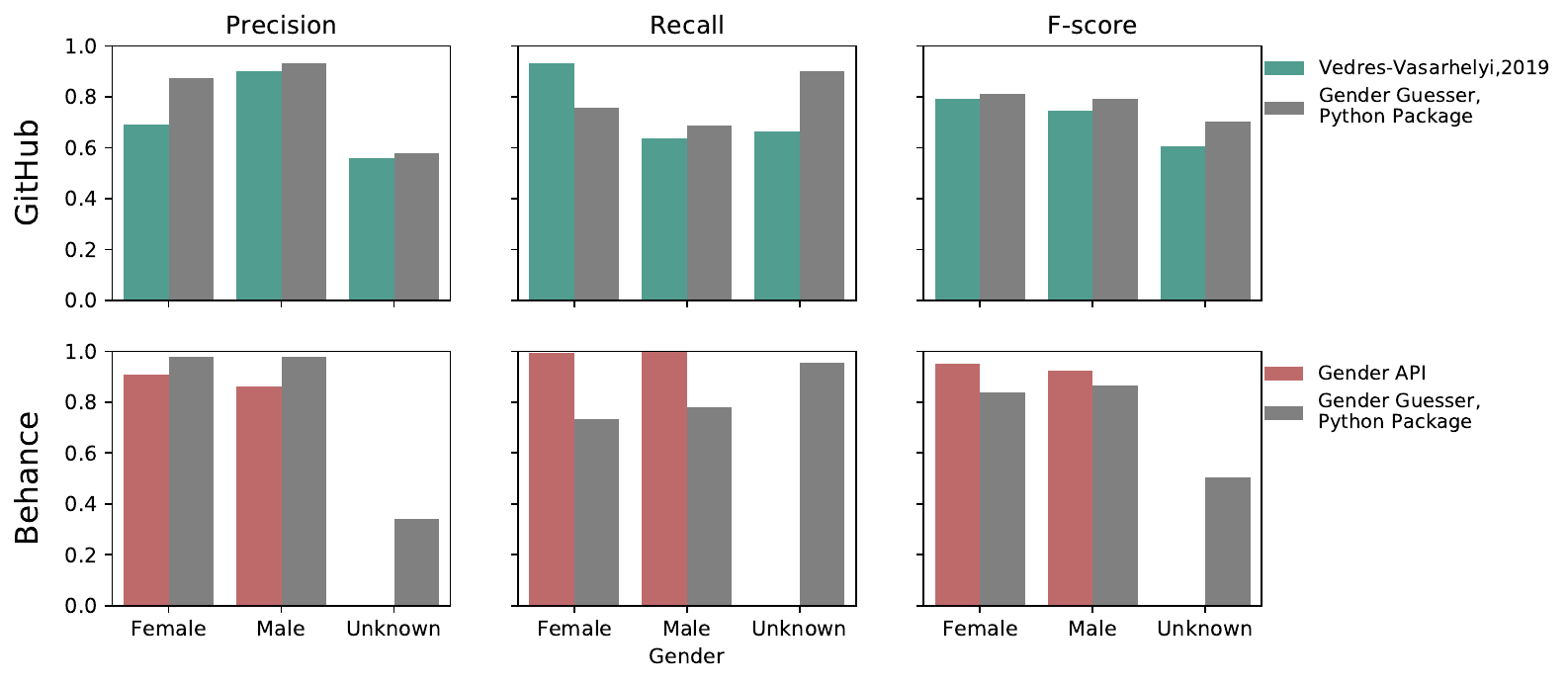}
\caption{ {\bf Gender Inferring Accuracy} Precision, Recall,  and F-score of the GitHub (Vedres-Vasarhelyi, 2019) and Behance (Gender API) gender inferring methods against the manually inferred baseline method and a commonly used alternative method (Gender Guesser Python Package). Among GitHub users, our method and the default Python package yielded very similar results, optimized for high male precision. The used method's relative strength is female-recall, and it's weakness is unknown-recall. The commercial Gender API used to infer the gender of Behance users resulted in higher overall precision, recall, and f-score compared to the default python package. It is important to note that this dataset officially did not include unknown-gendered users, although we found 45 (11\%) accounts which belong to companies, therefore, their gender could not be inferred.}
\label{fig:gen_inf}
\end{figure*}

In order to estimate the accuracy of these two gender inferring methods, we took a sample of 200 users for each gender category (female, male, unknown) from datasets, and inferred their gender manually. We compared our classification with gender inferring methods presented above and also added a third commonly used gender inferring method available as a ready-to-use Python package (Gender Guesser\footnote{https://pypi.org/project/gender-guesser/}). We found that among GitHub users our method and the default Python package yielded very similar results, optimized for high male precision. The commercial Gender API used to infer the gender of Behance users resulted in higher overall precision, recall, and f-score compared to the default Python package. (See precision, recall, and F-sore of each algorithm by gender at section Supplamantary Information Figure \ref{fig:gen_inf}). To validate the robustness of our results, we run all of our statistical models on datasets with varying levels of gender bias, that we introduced by swapping 5\%, 10\%, 25\% of the user's gender between male and female.

Finally, to fix the unbalanced nature of our data with regard to gender, we took a biased sample with 10,000 users of each gender (male, female) from the GitHub users and 6,000 each from Behance. We replicated our analysis on five samples; results in the main text are presented based on sample 1 (results for further samples are in our Supplementary Information).

\subsection*{Identifying specializations}

\begin{figure*}[ht]
\centering
\includegraphics[width=.95\linewidth]{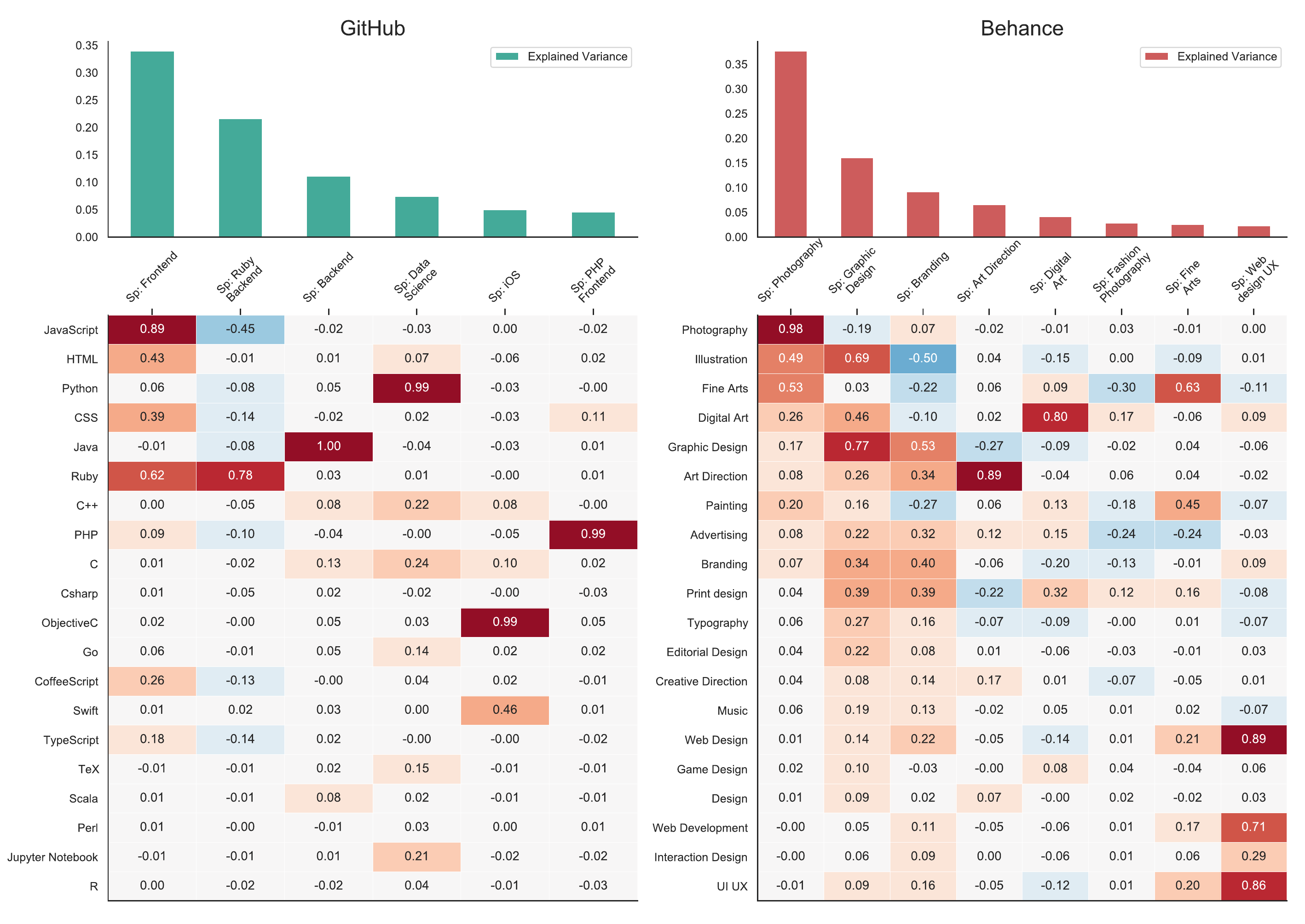}
\caption{\bf{Specializations on GitHub and Behance}  We used Scipy’s PCA.decomposition package with Varimax Rotation to identify independent factors \citep{Pedregosa2012}. Bar charts show the explained variance of each factor. The correlation matrices show the “importance” and the sign of the relationship of the language/design field in the component. On GitHub we identified 6 main specializations;  1) Frontend development, 2)  Developers using Ruby for backend development, 3)  Backend Development with high activity in Java, 4) Data Science, 5) iOS development, and  6) PHP enthusiastic with Frontend focus. On Behance our principal component analysis yield 8 main factors: 1) Photography, 2) Graphic Design, 3) Branding, 4) Art Direction, 5) Digital Art, 6) Fashion Photography 7) Fine Arts, and 8) Web design- UX.}
\label{fig:spec}
\end{figure*}

We used principal component analysis to identify specializations of users. In both datasets, we created field-specific count variables that measure the frequency at which a user worked with a given programming language on GitHub (e.g.: C, Java, Python), or the number of projects where the user indicated a given creative field (e.g.: painting, photography, copy-writing). For both platforms, we used the 20 most popular programming languages or design fields of those that appeared in at least 1000 projects. On GitHub we identified 6 main specializations;  1) Frontend development, 2)  Developers using Ruby for backend development, 3)  Backend Development with high activity in Java, 4) Data Science, 5) iOS (iPhone Operating System) development, and  6) PHP project with Frontend focus. In Behance our principal component analysis yielded eight main factors: 1) Photography, 2) Graphic Design, 3) Branding, 4) Art Direction, 5) Digital Art, 6) Fashion Photography, 7) Fine Arts and 8) Web design- UX. (See SI Figure \ref{fig:spec} for bar charts showing the explained variance of each factor and the correlation matrices showing the 'importance' and the sign of the relationship between the language / design fields in the resulting specialization.)

\subsection*{Femaleness}

We capture the gendered typicality of behavior as the probability of being female, given a pattern of activity. Specifically, we use Random Forest models to predict whether a user's inferred gender is female. Our features are variables that cover behavioral choices, such as type of engagement (creating  and modifying coding repositories, uploading design projects), specialization (programming languages or art categories) and networking (such as the number of followers). The resulting prediction score is {\it femaleness}, which quantifies the female typicality of creative behavior on a scale between 0 (most male-typical behavior) to 1 (most female-typical behavior). 

\begin{figure}[ht]
\centering
\includegraphics[width=1\linewidth]{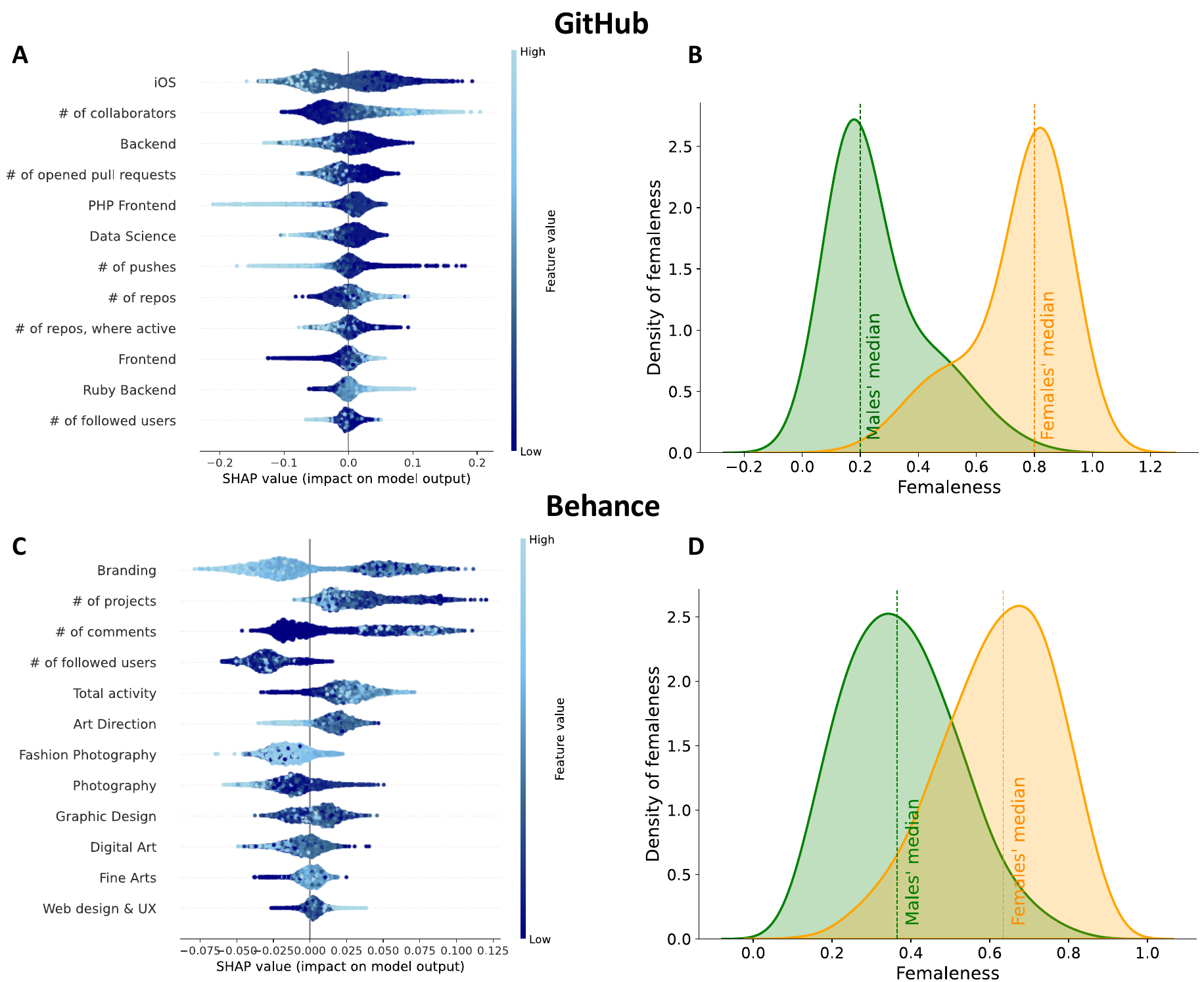}
\caption{A,C: Beeswarm plots of Femaleness. Each dot represents one data point, where the X axis is determined by the SHAP (SHapley Additive ExPlanations) value. The features are ordered by their relative importance on the Y axis. Color displays the original value of a feature - light colors indicate high values of the given feature, dark colors low. B,D: Distribution of Femaleness. Graphs represent the probability density of femaleness for males(green), females(orange) on GitHub (Panel B) and Behance (Panel D). Dashed lines indicate median femaleness by gender groups.}
\label{fig:femaleness}
\end{figure}

The GitHub Random Forest classification was moderately accurate ($AUC=0.64$), on Behance the accuracy was somewhat higher ($AUC=0.69$). A key strength of the Random Forest model is that it can capture nonlinear relationships between variables and enable intuitive ways to quantify the importance of variables \citep{Breiman2001, Strobl2008, parr2018beware}.

Panel A and C on Figure \ref{fig:femaleness} show how features impact the models' output (Femaleness) for each of our two cases. Dots represent users, and the horizontal axis shows SHAP (SHapley Additive ExPlanations) value, that estimates the contribution of a feature as the difference between expectation without the feature (the mean expected prediction) and the prediction with the feature for each given user. Color is used to display the original value of a feature - light colors indicate high values of the given feature, dark colors low. The features are ordered by their relative importance on the Y axis. For example, in the case of GitHub (on Panel A of Figure \ref{fig:femaleness}), the most important predictor of being female is developing software for the iOS mobile operating system (labeled "iOS", first row of Panel A), and the high values of iOS development (light blue dots in the first row of Panel A) predict a low probability that a user is female (as light blue dots are towards the lower, left hand side of the x-axis). This indicates that specializing in iOS development is more of a male-typical trait, rather than a female-typical one. Also, according to our model, a high number of collaborators and specialization in Ruby backend are more associated with being female, as the light blue dots in these rows are more towards the higher (right-hand) end of the x-axis. At Behance the most important predictor of being female is branding, low values of branding predict higher femaleness, indicating that it is a more masculine specialization. The high number of comments, specializing in fashion photography and Web design \& UX are the most feminine traits.

 Figures \ref{fig:femaleness} Panels B and D show the probability density of femaleness for males (green) and females (orange) on GitHub and Behance. The separation of developers by femaleness is more pronounced on GitHub; the median femaleness for females is $0.2$, and for males it is $0.8$. In Behance, the probability distributions and medians are closer to each other ($0.37$ and $0.63$ for males and females respectively). The distributions on Panels B and D indicate that behavioral pattern does differ by gender, although the distributions of femaleness for males and females do overlap. In other words, while females are often high on femaleness, we do find several females with low femaleness: male typical behavior.  

\subsection*{Models}

Table \ref{tab:vars} describes the variables used in modeling the impact of direct and indirect discrimination on three outcomes: {\it attention}, {\it  success} and {\it survival}. 

\begin{table*}[ht]
\centering
\begin{tabular}{lll}
\toprule
           & GitHub      & Behance    \\
\midrule
Attention  & \# of followers                                                                                                                 & \# of followers                                                                                                  \\
Success    & \# of stars on own repositories                                                                                                                     & \# of appreciations on own designs                                                                                              \\
Survival   & Activity one year after data collection                                                                                         & Activity one year after data collection                                                                          \\

Tenure     & Years since registration                                                                                                        & Years since registration                                                                                         \\
Gender & Inferred from nickname, email  or full name & Inferred from a user’s name\\
Activity   & \begin{tabular}[c]{@{}l@{}}\# of pushes, \# of own repos, \\ \# of repos, where active, \# of opened pull requests\end{tabular} & \begin{tabular}[c]{@{}l@{}}\# of projects, \# of comments, \\ total activity (views, appreciations)\end{tabular} \\
Networking & \# of collaborators, \# of of users followed                                                                                    & \# of of users followed                                                                                        \\
Fields & Programming languages used in projects & Creative fields designs labelled \\
\bottomrule
\end{tabular}
\caption{Variables computed for GitHub and Behance users.}
\label{tab:vars}
\end{table*}

We measure attention by the number of followers users have, which is available on both platforms. To become a follower, someone needs to be aware of a users' work and express willingness to keep updated about further works. Success is measured on GitHub by the number of stars on users' own repositories, and on Behance, by the number of "appreciations" (likes) on users' own designs. Success is more than merely attention, as it indicates expressed appreciation of quality towards a given piece of work from a user. To quantify survival, we revisited both platforms 365 days after data collection closed and checked whether the user had any additional activity in that 365-day interval. If the user did not leave any trace of activity within this one-year time window, we marked the user as inactive, otherwise we marked the user as a survivor.

Because attention and success are considerably skewed to the right, we apply a logarithmic transformation on the number of followers, number of stars, and project appreciations. Thus, $log(attention+1)$ and $log(success+1)$ are estimated with linear models. For the estimation of $survival$, since it is a binary variable (users who had activity marked with $1$, while dropped outs are marked as $0$), we used a logistic regression model.

We enter the same set of control variables in each model, corresponding to relevant alternative explanations for gender differences in outcomes. Due to higher work-family conflict and societal expectations, women generally have less time to maintain their professional presence; therefore, the level of activity might benefit men more than women\citep{Ahuja2002}. Men are more likely to join online portfolio sites earlier \citep{putzke2014cross}, and users with a longer tenure are more likely to build larger audiences and accumulate more visibility (attention) and success \citep{badashian2016measuring, jiang2017and}. Thus, we control for {\it  tenure} (number of years since registration) and {\it total activity} (Number of repositories or projects, and total activity on sites).

\begin{figure}[ht]
\centering
\includegraphics[width=1\linewidth]{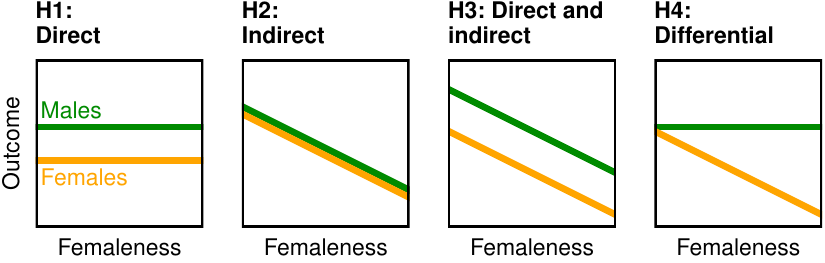}
\caption{Hypotheses regarding combinations of direct and indirect discrimination. Lines shows hypothetical marginal prediction of outcomes by gender category. Y axis is the resulting prediction of an outcome, X axis is femaleness, color indicates gender.}
\label{fig:hypothesis}
\end{figure}

We specified statistical models to identify the impact of both direct and indirect discrimination on outcomes. Therefore, our key variables, gender (binary, $1=Female$, $0=Male$) and femaleness, are entered to the models separately, and also with their interaction.  Figure \ref{fig:hypothesis} illustrates our hypotheses, separating the impact of direct discrimination by categorical gender (indicated by color) and indirect discrimination by gender typicality (femaleness, on the x axis) on outcomes. If individual outcomes were only impacted by direct discrimination (Figure \ref{fig:hypothesis}, Panel H1), only categorical gender would be a significant predictor in our models, without any slope for femaleness. In the inverse case, with only indirect discrimination (Figure \ref{fig:hypothesis}, Panel H2), femaleness would be a significant predictor in the models with a significant slope, without difference between the two gender groups (equally impacting both males and females). It is likely that outcomes are influenced by a combination of direct and indirect discrimination. When there is direct discrimination, the prediction lines will have significantly different intercepts by gender, and there will also be a significant slope by indirect discrimination that impacts male and female users in the same way (Figure \ref{fig:hypothesis}, Panel H3). Lastly, it is also possible that indirect discrimination will have a differing impact by gender, such that, for example, female users will be penalized when they follow female-typical behavior, but male users will not experience the same indirect discrimination (Figure \ref{fig:hypothesis}, H4). In such a case, the interaction term between direct and indirect discrimination will be significant.

\section*{Results}
We found that there is a significant baseline gender difference in attention, success, and survival on both platforms, with the exception of survival on Behance. The Mann-Whitney U tests (MW) revealed that men have a significantly higher number of followers {\small (GitHub: $IQR_{male}=[1,12]$, $IQR_{female}=[0,10]$ , $MW_{p}=0.000$ -- not significant via OLS, Behance: $IQR_{male}=[41;928]$, $IQR_{female}=[25,401]$, $MW_{p}=0.000$ )}, and  are more successful {\small (number of stars on GitHub $IQR_{male}=[0,1]$, $IQR_{female}=[0,0]$, $MW_{p}=0.000$, number of project appreciations on Behance $IQR_{male}=[66,2264]$, $IQR_{female}=[47,980]$, $MW_{p}=0.000$)} on both platforms. Men have a lower average dropout rate on GitHub {\small (average dropout $Avg_{male}=0.93$, $Avg_{female}=0.88$, $MW_{p}=0.000$)} but a higher one on Behance {\small ($Avg_{male}=0.45$, $Avg_{female}=0.417$ $MW_{p}=0.000$)}. (See SI Table \ref{tab:mw} for Mann-Whitney test results)

\begin{figure}[ht]
\centering
\includegraphics[width=1\linewidth]{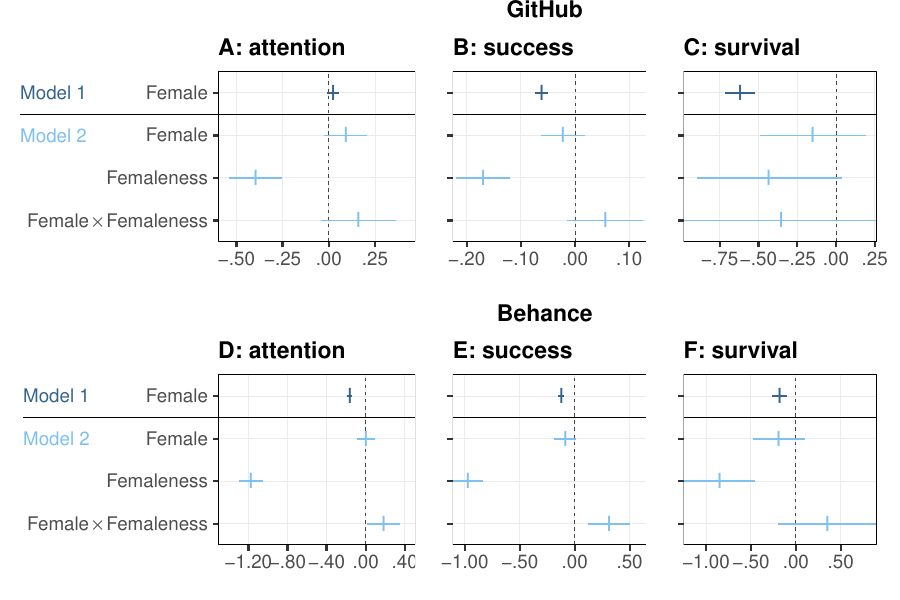}
\caption{Point estimates of outcomes, with 95 percent confidence intervals, for variables related to gender. {\it Attention} and {\it success} shows coefficients from Linear Models predicting success (the log. number of stars received, log. number of project appreciations), while {\it survival} shows odds ratios from logit models predicting survival over a one year period following our data collection.}
\label{fig:pointest}
\end{figure}

Still considering the gross difference between gender categories (without separating direct and indirect discrimination), but also entering controls for activity level, tenure, and fields, we still see a baseline categorical difference for gender in most outcomes. All variables are measured on the 0–1 scale, making estimates comparable. Figure \ref{fig:pointest} Model 1 for each of the six panels from A to F shows the relative difference for female developers in all outcomes and platforms, once we take controls into account. With the exception of attention on GitHub, where there is no significant difference (p =.186), all results show a significant female disadvantage (all $p=.000$). 

Figure \ref{fig:pointest}, Model 2. (across all panels from A to F) shows point estimates after including femaleness in the models. After introducing indirect discrimination, the gender category in itself (being female) is no longer significant. (This only means that at the zero value of femaleness, that is fully male typical behavior, there is no difference between gender categories in outcomes.) However, femaleness is a significant negative predictor of outcomes in all cases, except for survival on GitHub. This indicates that indirect discrimination is a significant predictor of differential results between men and women in attention, success on both platforms, and in survival on Behance. Our models also indicate that it is only on Behance in the case of attention and success (Figure \ref{fig:pointest} panels D and E) where femaleness impacts men and women significantly differently: Women with high femaleness are predicted to receive more attention and have more project appreciations, suggesting that men are more penalized for exhibiting highly female-like behavior. 

Our results are consistent across all 5 samples. Femaleness remains significant in gender-swapped datasets with error rates 5\% in 70- 87\% of the cases (out of 100 reruns) and with 10\% error rates, 55- 86\% on GitHub and 81- 90\% on Behance. Simulations with 25\% gender swapping are less consistent with significant cases; less than 50\% remains significant. (See Model Table \ref{tab:Model_github} and \ref{tab:Model_behance} for 5 samples and gender-swapped simulations at SI)

\begin{figure}[ht]
\centering
\includegraphics[width=1\linewidth]{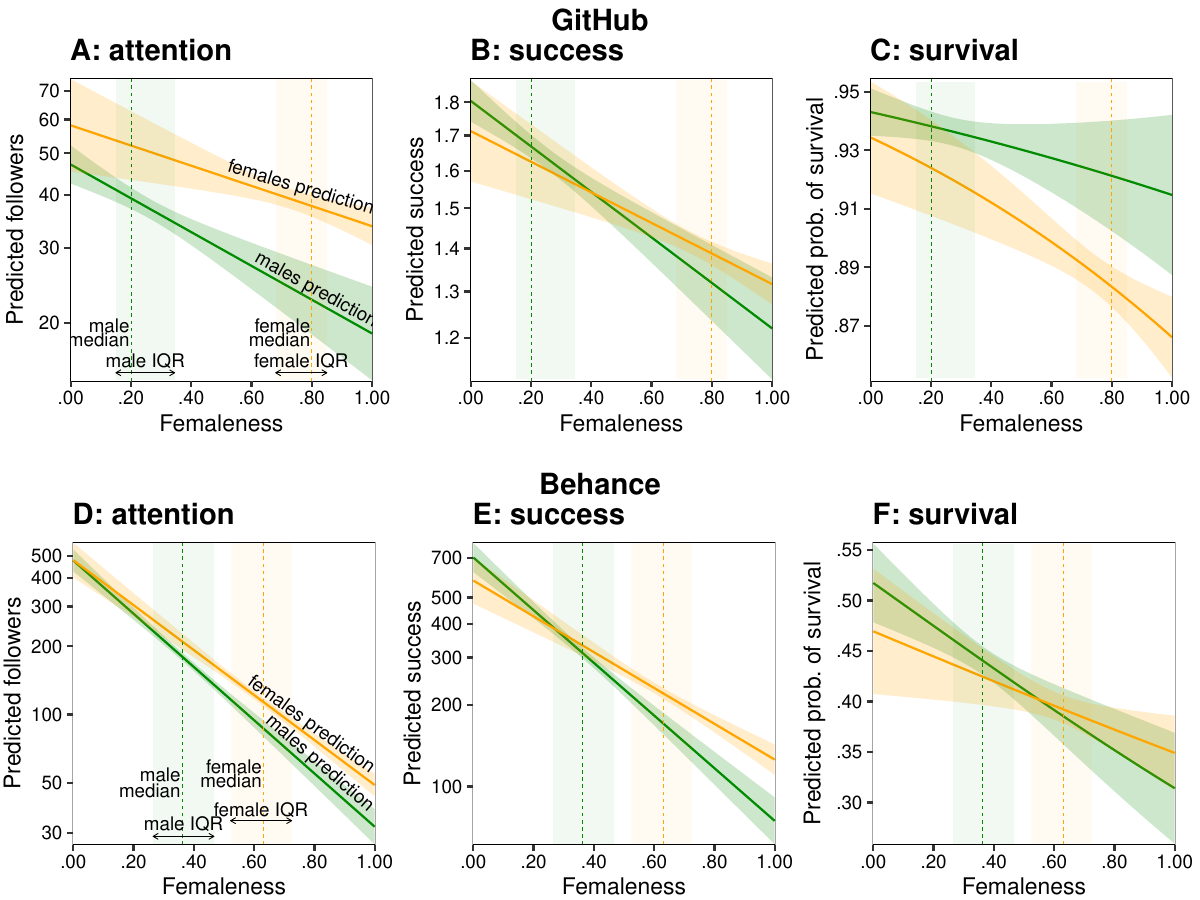}
\caption{Marginal predictions of outcomes from model 2 from Fig. \ref{fig:pointest}, with fixing all other variables at their means. Vertical dashed lines indicate medians of femaleness, and shaded vertical bars show the interquartile range (IQR)}
\label{fig:margins}
\end{figure}

Figure \ref{fig:margins} shows the predicted values of attention (1st column), success (2nd column), and survival (3rd column) along the range of femaleness by gender categories on GitHub (1st row) and Behance (2nd row). Although the negative impact of femaleness put both men and women in disadvantage in all models (negative slope), in some cases categorical gender impacts outcomes differently. In the case of attention, the difference between women (orange) and men (green) increases along the range of femaleness, predicting a higher level of attention for women users with highly female-like behavior than men. This trend holds for predicting the number of project appreciations on Behance, while the difference between female and male GitHub users' success by femaleness is insignificant. Categorical gender has no impact on the survival of Behance users, while female GitHub users are clearly in disadvantage compared to men.

Figure \ref{fig:margins} allows for testing our hypotheses described on Figure \ref{fig:hypothesis}. Shared across all panels of Figure \ref{fig:margins} is the presence of indirect discrimination. Unambiguous direct discrimination -- as consistent categorical disadvantage for women when the prediction of women never cross the prediction for men -- is only present in the case of survival on GitHub (Figure \ref{fig:margins} panel C).

To quantify female disadvantage, we take the prediction of the outcomes of men at their median femaleness and deduct the predicted value of women at their femaleness. On GitHub women have an attention gap of $1.62$ followers. Relative to the predicted number of followers of men, at men's median femaleness, it is a 4\% gap. The disadvantage is so small, because men suffer more from indirect discrimination than women (men lose $17$ followers between their and women's femaleness median, while women lose only $15$), and women have a direct gender advantage in attention ($13$ more followers than men at men's median). Women suffer a total attention disadvantage of 26\% relative to men on Behance. Although women have a direct gender advantage compared to men, it cannot compensate that they are more affected by indirect discrimination. 

On GitHub women have a total success disadvantage of 6\%, of which 90\% is due to indirect discrimination and 10\% due to direct. On Behance women have a 37\% total disadvantage compared to men's in success, which is entirely caused by indirect discrimination and mediated by 20\% by a direct female advantage.

In predicted survival, women suffer a total disadvantage of 6\% on GitHub, out of which 74\% is due to indirect discrimination and 26\% due to direct discrimination. The trends are similar on Behance with a total of 11\% of female disadvantage, composed of 60\% indirect and 40\% direct discrimination. (See Table SI.4. for prediction results at male and female femaleness medians by outcomes, and calculated direct and indirect discrimination in exact numbers and percentage)

\section{Conclusion}

Collaborative platforms show consistent indirect gender discrimination, while direct discrimination is only occasional and small. Our findings indicate that indirect gender discrimination -- rooted in career choices and online activity -- is present both on GitHub (a considerably male-dominated platform), and also on Behance (a platform with higher female presence). We found that the main cause of women's disadvantage in attention, success, and survival is largely due to indirect discrimination that varies between 60-90\% of total female disadvantage. 

Indirect discrimination negatively impacts both genders, furthermore, in the more gender-balanced design community, men are penalized even more for female-typical behavior, compared to women.  Although design careers are often considered more feminine than software development, and have a higher ratio of female professionals (28\%), masculine career choices have previously been associated with greater success \citep{wachs2017}. Although empirical studies suggest that shorter tenure and lower participation rates are the main factor behind lower wages and success in creative fields \citep{cook2018gender}, the case of Behance proves the opposite: in this case, higher participation of women does not automatically diminish indirect gender discrimination.

We found supporting evidence for the visibility paradox of women in technical fields. Women attract considerably more attention, but are often not recognized as experts \citep{Fernando2018}. Although femaleness is negatively related to attention on both fields, and women suffer from even higher indirect discrimination on Behance, they have a direct gender advantage in attention. However, female developers on GitHub are extremely visible, they enjoy almost a 9 times higher gender advantage compared to men. On Behance, the visibility gap caused by femaleness is only reduced by 45\% due to categorical gender. 

\begin{figure}[ht]
\centering
\includegraphics[width=.9\linewidth]{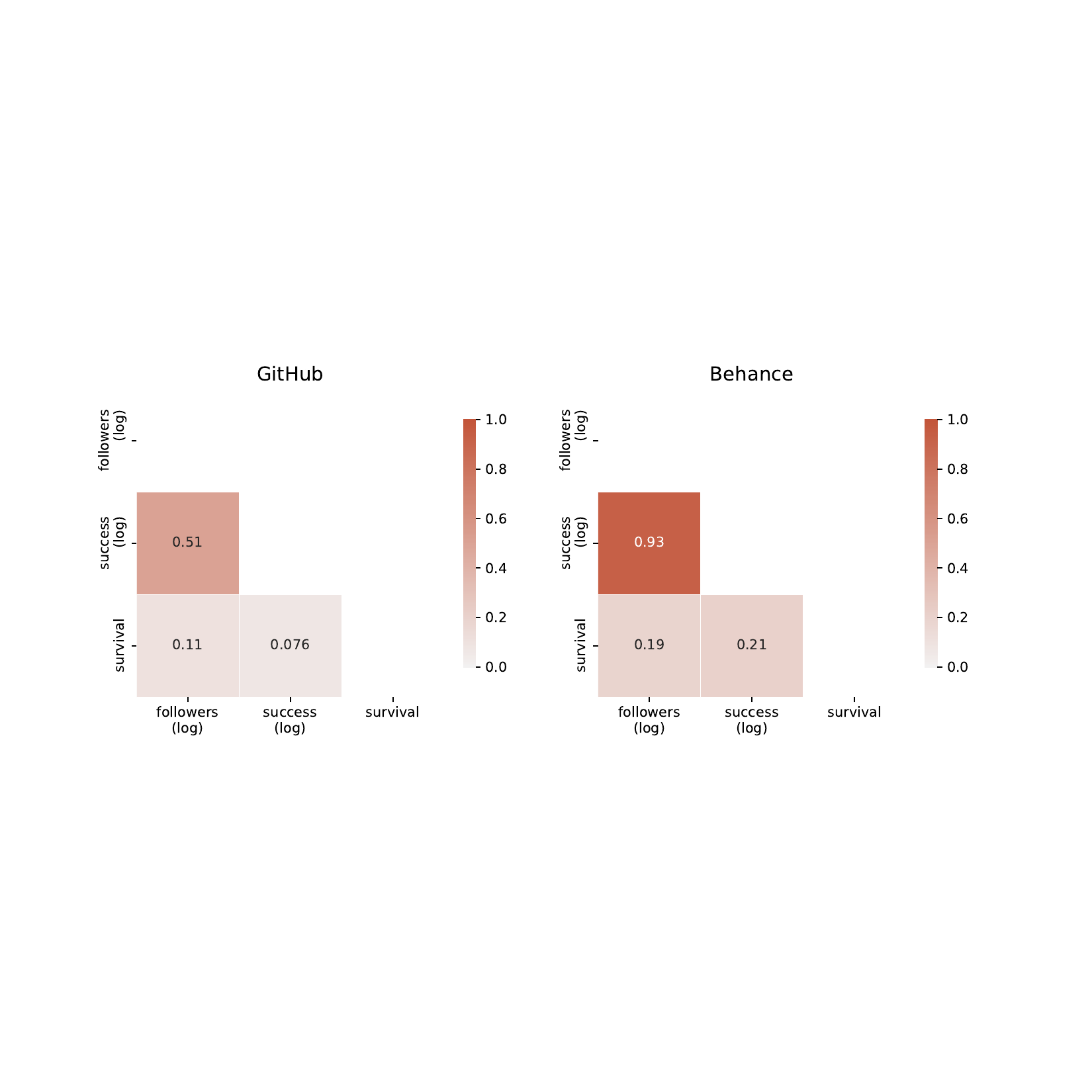}
\caption{ Pearson Correlation of outcome variables of Behance and GitHub}
\label{fig:gcorr}
\end{figure}

The higher attention that women attract on online collaborative platforms is a double-edge sword: On one hand, female professionals could use this increased visibility to build a larger audience and promote their work, which can eventually help them succeed through more visible role models \citep{eccles2016motivates, Kahn2017}. We cannot test the casual hypothesis that attention leads to success and survival (as our platform data do not offer the opportunity for quasi-experimental setups), but we do see evidence of association. Attention is highly correlated with success (($corr_{GitHub}=.51$, $p=.00$, $corr_{Behance}=.93$, $p=.00$). 
On the other hand, increased attention also has negative consequences for women. Women are more likely to be harassed online, and increased visibility could attract verbal violence as well \citep{yelin2020doing, moss2010disruptions, atir2018gender}, ultimately making women less likely to participate \citep{coffman2014evidence} and dropping out at higher rates. We see signs for this, as survival also correlates with visibility and success on both platforms, but the relationship is stronger at Behance. This can indicate that women on Behance could use their increased visibility to counteract indirect gender discrimination, gain more success, and eventually stay active on the platform (see SI Figure \ref{fig:gcorr} for correlation among outcome variables).

As platforms employ algorithmic tools and AI systems to manage users' activity, visibility and recommend new projects to collaborate, stereotypes rooted in behavior can have long-lasting consequences. Supposedly, 'gender-blind' self-learning algorithms trained on biased data will reproduce indirect gender bias in outcomes, and detect correlation patterns that can easily reveal users' gender\citep{gerards2021algorithmic}.

Since the launch of ChatGPT, large language models have become a key tool for asking programming-related questions and creating digital arts\citep{del2023large}. These models were trained on data from the Internet and have been shown to reproduce biases inherent in their data sources\citep{kotek2023gender}. If indirect discrimination is highly associated with success in online platforms, products created via such AI systems might prefer solutions and creative outputs generated by users with less female-like behavior. There are already signs that self-learning algorithms would propagate existing gender disparities in the labor market \citep{ali2019discrimination}, search engines \citep{vlasceanu2022propagation} and produce images in a sexist fashion \citep{bolukbasi2016man, zou2018ai}. As these models would not have the capacity to recognize and resist gender stereotypes baked into their training data sources, solutions built with them will carry over such stereotypes. As AI solutions become wide-spread, we fear that it will become almost impossible to detect and evade indirect gender discrimination. The only way to mitigate the impact of AI-amplified indirect discrimination would be to put into place a monitoring mechanism that constantly measures direct and indirect discrimination to alert the public to intervene against harms to underrepresented groups.  

\bibliographystyle{plain}
\bibliography{template}

\newpage
\section{Supplementary Information}

\begin{table}[ht]
\centering
\begin{tabular}{llllllll}
\toprule
                        &           & \multicolumn{1}{c}{\multirow{2}{*}{MW}} & \multicolumn{1}{c}{\multirow{2}{*}{P}} & \multicolumn{2}{c}{Female}  & \multicolumn{2}{c}{Male}    \\
                        &           & \multicolumn{1}{c}{}                    & \multicolumn{1}{c}{}                   & .25 quantile & .75 quantile & .25 quantile & .75 qunatile \\
                        \midrule
Behance                 & Attention & 21135593                                & 0,000                                  & 25           & 401          & 41           & 927.5        \\
                        & Success   & 20261832                                & 0.000                                  & 47           & 980          & 66           & 2264         \\
                        & Survival  & 17853000                                & 0.128                                  & 1            & 1            & 1            & 1            \\
                        \midrule
\multirow{3}{*}{GitHub} & Attention & 54302043                                & 0.000                                  & 0            & 10           & 1            & 12           \\
                        & Success   & 55770990                                & 0.000                                  & 0            & 0            & 0            & 1            \\
                        & Survival  & 52335000                                & 0.000                                  & 1            & 1            & 1            & 1           \\
                        \bottomrule

\end{tabular}
\caption{Mann-Whitney Statistics (MW),Significance (P) and Interquartile range of Attention, Success, Survival by gender}
\label{tab:mw}
\end{table}

\begin{sidewaystable}[!ht]
\begin{adjustbox}{width=\linewidth}
\begin{tabular}{crrrrrrrrrrrrrrrrrrrrrrrrrrrr}
\toprule
                           &                       & \multicolumn{3}{c}{Sample 1}                                                     & \multicolumn{3}{c}{Sample 2}                                                     & \multicolumn{3}{c}{Sample 3}                                                     & \multicolumn{3}{c}{Sample 4}                                                     & \multicolumn{3}{c}{Sample 5}                                                     & \multicolumn{4}{c}{5\%}                                                                                  & \multicolumn{4}{c}{10\%}                                                                                 & \multicolumn{4}{c}{25\%}                                                                                 \\
                           & Variable              & \multicolumn{1}{c}{Coef.} & \multicolumn{1}{c}{SE.} & \multicolumn{1}{c}{P} & \multicolumn{1}{c}{Coef.} & \multicolumn{1}{c}{SE.} & \multicolumn{1}{c}{P} & \multicolumn{1}{c}{Coef.} & \multicolumn{1}{c}{SE.} & \multicolumn{1}{c}{P} & \multicolumn{1}{c}{Coef.} & \multicolumn{1}{c}{SE.} & \multicolumn{1}{c}{P} & \multicolumn{1}{c}{Coef.} & \multicolumn{1}{c}{SE.} & \multicolumn{1}{c}{P} & \multicolumn{1}{c}{Avg.} & \multicolumn{1}{c}{Min} & \multicolumn{1}{c}{Max} & \multicolumn{1}{c}{Sign.} & \multicolumn{1}{c}{Avg.} & \multicolumn{1}{c}{Min} & \multicolumn{1}{c}{Max} & \multicolumn{1}{c}{Sign.} & \multicolumn{1}{c}{Avg.} & \multicolumn{1}{c}{Min} & \multicolumn{1}{c}{Max} & \multicolumn{1}{c}{Sign.} \\
\midrule
\multirow{12}{*}{\rotatebox[origin=c]{90}{Attention}} & Female                & 0.036                     & 0.025                        & 0.147                 & 0.038                     & 0.026                        & 0.144                 & 0.063                     & 0.026                        & 0.015                 & 0.028                     & 0.025                        & 0.272                 & 0.043                     & 0.026                        & 0.088                 & 0.025                    & -0.026                  & 0.095                   & 20\%                      & 0.014                    & -0.028                  & 0.081                   & 9\%                       & 0.012                    & -0.024                  & 0.060                   & 6\%                       \\
                           & Femaleness            & -0.165                    & 0.033                        & 0.000                 & -0.135                    & 0.032                        & 0.000                 & -0.185                    & 0.033                        & 0.000                 & -0.174                    & 0.032                        & 0.000                 & -0.105                    & 0.032                        & 0.001                 & -0.078                   & -0.151                  & 0.014                   & 70\%                      & -0.059                   & -0.113                  & 0.014                   & 55\%                      & -0.022                   & -0.090                  & 0.045                   & 19\%                      \\
                           & \begin{tabular}[c]{@{}r@{}}Female:\\ Femaleness\end{tabular}     & 0.074                     & 0.045                        & 0.101                 & 0.053                     & 0.045                        & 0.236                 & 0.045                     & 0.045                        & 0.323                 & 0.075                     & 0.044                        & 0.093                 & 0.022                     & 0.045                        & 0.618                 & 0.024                    & -0.094                  & 0.126                   & 13\%                      & 0.025                    & -0.095                  & 0.103                   & 12\%                      & 0.004                    & -0.082                  & 0.067                   & 5\%                       \\
                           & \begin{tabular}[c]{@{}r@{}}N. Own \\ Repos (log)\end{tabular}  & -0.067                    & 0.006                        & 0.000                 & -0.066                    & 0.006                        & 0.000                 & -0.055                    & 0.006                        & 0.000                 & -0.067                    & 0.006                        & 0.000                 & -0.059                    & 0.006                        & 0.000                 & -0.055                   & -0.056                  & -0.054                  & 90\%                      & -0.056                   & -0.056                  & -0.055                  & 90\%                      & -0.056                   & -0.057                  & -0.056                  & 90\%                      \\
                            & \begin{tabular}[c]{@{}r@{}}N. Active \\ Repos (log)\end{tabular} & 0.989                     & 0.018                        & 0.000                 & 0.955                     & 0.018                        & 0.000                 & 0.960                     & 0.018                        & 0.000                 & 0.995                     & 0.018                        & 0.000                 & 0.954                     & 0.018                        & 0.000                 & 0.964                    & 0.961                   & 0.967                   & 90\%                      & 0.965                    & 0.962                   & 0.967                   & 90\%                      & 0.966                    & 0.964                   & 0.967                   & 90\%                      \\
                           & Tenure                & 0.167                     & 0.004                        & 0.000                 & 0.169                     & 0.004                        & 0.000                 & 0.169                     & 0.004                        & 0.000                 & 0.171                     & 0.004                        & 0.000                 & 0.170                     & 0.004                        & 0.000                 & 0.171                    & 0.170                   & 0.172                   & 90\%                      & 0.171                    & 0.170                   & 0.172                   & 90\%                      & 0.172                    & 0.171                   & 0.172                   & 90\%                      \\
                           & Intercept             & -0.878                    & 0.024                        & 0.000                 & -0.847                    & 0.024                        & 0.000                 & -0.872                    & 0.024                        & 0.000                 & -0.885                    & 0.024                        & 0.000                 & -0.877                    & 0.024                        & 0.000                 & -0.907                   & -0.937                  & -0.883                  & 90\%                      & -0.912                   & -0.938                  & -0.894                  & 90\%                      & -0.924                   & -0.951                  & -0.895                  & 90\%                      \\
                           & N                     &                           &                              & 20000                 &                           &                              & 20000                 &                           &                              & 20000                 &                           &                              & 20000                 &                           &                              & 20000                 &                          &                         &                         &                           &                          &                         &                         &                           &                          &                         &                         &                           \\
                           & R2                    &                           &                              & 0.259                 &                           &                              & 0.258                 &                           &                              & 0.263                 &                           &                              & 0.265                 &                           &                              & 0.254                 &                          &                         &                         &                           &                          &                         &                         &                           &                          &                         &                         &                           \\
\midrule
\multirow{12}{*}{\rotatebox[origin=c]{90}{Success}}   & Female                & -0.001                    & 0.020                        & 0.977                 & -0.021                    & 0.021                        & 0.307                 & -0.018                    & 0.021                        & 0.400                 & -0.009                    & 0.021                        & 0.664                 & -0.016                    & 0.020                        & 0.432                 & -0.029                   & -0.066                  & 0.006                   & 37\%                      & -0.046                   & -0.085                  & -0.011                  & 76\%                      & -0.058                   & -0.080                  & -0.021                  & 89\%                      \\
                           & Femaleness            & -0.142                    & 0.026                        & 0.000                 & -0.135                    & 0.025                        & 0.000                 & -0.165                    & 0.026                        & 0.000                 & -0.213                    & 0.026                        & 0.000                 & -0.175                    & 0.026                        & 0.000                 & -0.103                   & -0.146                  & -0.032                  & 87\%                      & -0.078                   & -0.116                  & -0.023                  & 86\%                      & -0.034                   & -0.082                  & 0.014                   & 49\%                      \\
                           & \begin{tabular}[c]{@{}r@{}}Female:\\ Femaleness\end{tabular}   & 0.009                     & 0.036                        & 0.802                 & 0.030                     & 0.036                        & 0.403                 & 0.045                     & 0.036                        & 0.220                 & 0.052                     & 0.036                        & 0.147                 & 0.052                     & 0.036                        & 0.147                 & 0.015                    & -0.054                  & 0.080                   & 7\%                       & 0.018                    & -0.056                  & 0.084                   & 6\%                       & 0.006                    & -0.072                  & 0.054                   & 4\%                       \\
                           & \begin{tabular}[c]{@{}r@{}}N. Own \\ Repos (log)\end{tabular}    & -0.022                    & 0.005                        & 0.000                 & -0.007                    & 0.005                        & 0.144                 & -0.015                    & 0.005                        & 0.001                 & -0.009                    & 0.005                        & 0.051                 & -0.009                    & 0.005                        & 0.063                 & -0.015                   & -0.016                  & -0.014                  & 90\%                      & -0.015                   & -0.016                  & -0.015                  & 90\%                      & -0.016                   & -0.016                  & -0.015                  & 90\%                      \\
                            & \begin{tabular}[c]{@{}r@{}}N. Active \\ Repos (log)\end{tabular} & 0.585                     & 0.014                        & 0.000                 & 0.534                     & 0.014                        & 0.000                 & 0.557                     & 0.014                        & 0.000                 & 0.568                     & 0.015                        & 0.000                 & 0.540                     & 0.014                        & 0.000                 & 0.559                    & 0.556                   & 0.563                   & 90\%                      & 0.560                    & 0.558                   & 0.563                   & 90\%                      & 0.562                    & 0.559                   & 0.564                   & 90\%                      \\
                           & Tenure                & 0.086                     & 0.003                        & 0.000                 & 0.082                     & 0.003                        & 0.000                 & 0.082                     & 0.003                        & 0.000                 & 0.086                     & 0.003                        & 0.000                 & 0.081                     & 0.003                        & 0.000                 & 0.082                    & 0.082                   & 0.083                   & 90\%                      & 0.083                    & 0.082                   & 0.084                   & 90\%                      & 0.084                    & 0.083                   & 0.084                   & 90\%                      \\
                           & Intercept             & -0.720                    & 0.019                        & 0.000                 & -0.683                    & 0.019                        & 0.000                 & -0.687                    & 0.019                        & 0.000                 & -0.700                    & 0.019                        & 0.000                 & -0.677                    & 0.020                        & 0.000                 & -0.705                   & -0.729                  & -0.691                  & 90\%                      & -0.713                   & -0.733                  & -0.700                  & 90\%                      & -0.726                   & -0.746                  & -0.705                  & 90\%                      \\
                           & N                     &                           &                              & 20000                 &                           &                              & 20000                 &                           &                              & 20000                 &                           &                              & 20000                 &                           &                              & 20000                 &                          &                         &                         &                           &                          &                         &                         &                           &                          &                         &                         &                           \\
                           & R2                    &                           &                              & 0.170                 &                           &                              & 0.163                 &                           &                              & 0.164                 &                           &                              & 0.173                 &                           &                              & 0.159                 &                          &                         &                         &                           &                          &                         &                         &                           &                          &                         &                         &                           \\
\midrule
\multirow{13}{*}{\rotatebox[origin=c]{90}{Survival}} & Female                & -0.112                    & 0.179                        & 0.530                 & -0.340                    & 0.178                        & 0.056                 & -0.059                    & 0.185                        & 0.748                 & -0.175                    & 0.178                        & 0.326                 & -0.372                    & 0.179                        & 0.037                 & -0.029                   & -0.066                  & 0.006                   & 37\%                      & -0.046                   & -0.085                  & -0.011                  & 76                        & -0.058                   & -0.080                  & -0.021                  & 89\%                      \\
                           & Femaleness            & -0.949                    & 0.240                        & 0.000                 & -0.340                    & 0.252                        & 0.177                 & -0.764                    & 0.245                        & 0.002                 & -0.661                    & 0.248                        & 0.008                 & -0.789                    & 0.244                        & 0.001                 & -0.103                   & -0.146                  & -0.032                  & 87\%                      & -0.078                   & -0.116                  & -0.023                  & 86                        & -0.034                   & -0.082                  & 0.014                   & 49\%                      \\
                           & \begin{tabular}[c]{@{}r@{}}Female:\\ Femaleness\end{tabular}    & -0.104                    & 0.319                        & 0.744                 & -0.335                    & 0.325                        & 0.304                 & -0.361                    & 0.326                        & 0.267                 & -0.340                    & 0.322                        & 0.290                 & 0.059                     & 0.320                        & 0.854                 & 0.015                    & -0.054                  & 0.080                   & 7\%                       & 0.018                    & -0.056                  & 0.084                   & 6                         & 0.006                    & -0.072                  & 0.054                   & 4\%                       \\
                           & \begin{tabular}[c]{@{}r@{}}N. Own \\ Repos (log)\end{tabular}   & -0.208                    & 0.047                        & 0.000                 & -0.271                    & 0.049                        & 0.000                 & -0.207                    & 0.047                        & 0.000                 & -0.163                    & 0.046                        & 0.000                 & -0.263                    & 0.049                        & 0.000                 & -0.015                   & -0.016                  & -0.014                  & 90\%                      & -0.015                   & -0.016                  & -0.015                  & 90                        & -0.016                   & -0.016                  & -0.015                  & 90\%                      \\
                           & \begin{tabular}[c]{@{}r@{}}N. Active \\ Repos (log)\end{tabular} & 2.168                     & 0.157                        & 0.000                 & 2.530                     & 0.161                        & 0.000                 & 2.157                     & 0.155                        & 0.000                 & 2.059                     & 0.155                        & 0.000                 & 2.540                     & 0.164                        & 0.000                 & 0.559                    & 0.556                   & 0.563                   & 90\%                      & 0.560                    & 0.558                   & 0.563                   & 90                        & 0.562                    & 0.559                   & 0.564                   & 90\%                      \\
                           & Tenure                & -0.334                    & 0.028                        & 0.000                 & -0.335                    & 0.028                        & 0.000                 & -0.329                    & 0.028                        & 0.000                 & -0.309                    & 0.028                        & 0.000                 & -0.333                    & 0.028                        & 0.000                 & 0.082                    & 0.082                   & 0.083                   & 90\%                      & 0.083                    & 0.082                   & 0.084                   & 90                        & 0.084                    & 0.083                   & 0.084                   & 90\%                      \\
                           & Intercept             & 1.144                     & 0.192                        & 0.000                 & 0.728                     & 0.193                        & 0.000                 & 1.123                     & 0.193                        & 0.000                 & 1.117                     & 0.191                        & 0.000                 & 0.796                     & 0.200                        & 0.000                 & -0.705                   & -0.729                  & -0.691                  & 90\%                      & -0.713                   & -0.733                  & -0.700                  & 90                        & -0.726                   & -0.746                  & -0.705                  & 90\%                      \\
                           & N                     &                           &                              & 20000                 &                           &                              & 20000                 &                           &                              & 20000                 &                           &                              & 20000                 &                           &                              & 20000                 &                          &                         &                         &                           &                          &                         &                         &                           &                          &                         &                         &                           \\
                           & AIC                   &                           &                              & 20792                 &                           &                              & 20519                 &                           &                              & 20879                 &                           &                              & 21651                 &                           &                              & 20760                 &                          &                         &                         &                           &                          &                         &                         &                           &                          &                         &                         &                           \\
                           & BIC                   &                           &                              & 20847                 &                           &                              & 20575                 &                           &                              & 20934                 &                           &                              & 21706                 &                           &                              & 20815                 &                          &                         &                         &                           &                          &                         &                         &                           &                          &                         &                         &   \\
                           \bottomrule
\end{tabular}
\end{adjustbox}
\caption{GitHub models. OLS Regression Model tables predicting Log(Attention), Log(Success), and logistic regression models predicting staying active on GitHub one year after data collection. All variables are normalized between 0 and 1. Samples indicate independent samples containing 10,000 female and male GitHub users. Robustness results are run on 100 5\%, 10\%, and 25\% randomly gender-swapped datasets. Avg. shows the average coefficients of the models, Min and Max indicate the lowest and highest values of 100 runs, and Sign. shows the ratio of significant coefficients.}
\label{tab:Model_github}
\end{sidewaystable}\textbf{}

\begin{sidewaystable}[ht]
\begin{adjustbox}{width=\linewidth}
\begin{tabular}{crrrrrrrrrrrrrrrrrrrrrrrrrrrr}
\toprule
                           &                       & \multicolumn{3}{c}{Sample 1}                                                     & \multicolumn{3}{c}{Sample 2}                                                     & \multicolumn{3}{c}{Sample 3}                                                     & \multicolumn{3}{c}{Sample 4}                                                     & \multicolumn{3}{c}{Sample 5}                                                     & \multicolumn{4}{c}{5\%}                                                                                  & \multicolumn{4}{c}{10\%}                                                                                 & \multicolumn{4}{c}{25\%}                                                                                 \\
                           & Variable              & \multicolumn{1}{c}{Coef.} & \multicolumn{1}{c}{SE.} & \multicolumn{1}{c}{P} & \multicolumn{1}{c}{Coef.} & \multicolumn{1}{c}{SE.} & \multicolumn{1}{c}{P} & \multicolumn{1}{c}{Coef.} & \multicolumn{1}{c}{SE.} & \multicolumn{1}{c}{P} & \multicolumn{1}{c}{Coef.} & \multicolumn{1}{c}{SE.} & \multicolumn{1}{c}{P} & \multicolumn{1}{c}{Coef.} & \multicolumn{1}{c}{SE.} & \multicolumn{1}{c}{P} & \multicolumn{1}{c}{Avg.} & \multicolumn{1}{c}{Min} & \multicolumn{1}{c}{Max} & \multicolumn{1}{c}{Sign.} & \multicolumn{1}{c}{Avg.} & \multicolumn{1}{c}{Min} & \multicolumn{1}{c}{Max} & \multicolumn{1}{c}{Sign.} & \multicolumn{1}{c}{Avg.} & \multicolumn{1}{c}{Min} & \multicolumn{1}{c}{Max} & \multicolumn{1}{c}{Sign.} \\
\midrule
\multirow{12}{*}{\rotatebox[origin=c]{90}{Attention}} & Female                                                          & -0.040                    & 0.045                   & 0.374                 & -0.018                    & 0.046                   & 0.694                 & -0.063                    & 0.046                   & 0.170                 & 0.011                     & 0.046                   & 0.814                 & -0.046                    & 0.046                   & 0.312                 & -0.059                   & -0.148                  & 0.051                   & 16\%                      & -0.108                   & -0.201                  & 0.049                   & 57\%                      & -0.179                   & -0.312                  & 0.003                   & 76\%                      \\
\multicolumn{1}{c}{}                           & Femaleness                                                      & -1.023                    & 0.061                   & 0.000                 & -1.019                    & 0.059                   & 0.000                 & -1.050                    & 0.061                   & 0.000                 & -1.126                    & 0.060                   & 0.000                 & -1.113                    & 0.061                   & 0.000                 & -1.021                   & -1.171                  & -0.808                  & 90\%                      & -0.923                   & -1.149                  & -0.688                  & 90\%                      & -0.279                   & -0.964                  & 0.211                   & 76\%                      \\
\multicolumn{1}{c}{}                           & \begin{tabular}[c]{@{}r@{}}Female:\\ Femaleness\end{tabular}    & 0.191                     & 0.084                   & 0.023                 & 0.154                     & 0.084                   & 0.066                 & 0.236                     & 0.085                   & 0.005                 & 0.146                     & 0.085                   & 0.086                 & 0.237                     & 0.085                   & 0.005                 & 0.164                    & -0.075                  & 0.289                   & 39\%                      & 0.158                    & -0.091                  & 0.322                   & 32\%                      & 0.052                    & -0.328                  & 0.327                   & 13\%                      \\
\multicolumn{1}{c}{}                           & Tenure                                                          & 0.154                     & 0.003                   & 0.000                 & 0.154                     & 0.003                   & 0.000                 & 0.154                     & 0.003                   & 0.000                 & 0.153                     & 0.003                   & 0.000                 & 0.155                     & 0.003                   & 0.000                 & 0.651                    & 0.628                   & 0.683                   & 90\%                      & 0.634                    & 0.603                   & 0.658                   & 90\%                      & 0.600                    & 0.590                   & 0.621                   & 90\%                      \\
\multicolumn{1}{c}{}                           & \begin{tabular}[c]{@{}r@{}}N. projects \\ (log)\end{tabular}    & 0.655                     & 0.020                   & 0.000                 & 0.656                     & 0.020                   & 0.000                 & 0.655                     & 0.020                   & 0.000                 & 0.670                     & 0.020                   & 0.000                 & 0.649                     & 0.020                   & 0.000                 & 0.155                    & 0.153                   & 0.157                   & 90\%                      & 0.157                    & 0.153                   & 0.159                   & 90\%                      & 0.161                    & 0.158                   & 0.162                   & 90\%                      \\
\multicolumn{1}{c}{}                           & \begin{tabular}[c]{@{}r@{}}Total activity\\  (log)\end{tabular} & 0.649                     & 0.038                   & 0.000                 & 0.655                     & 0.038                   & 0.000                 & 0.677                     & 0.038                   & 0.000                 & 0.623                     & 0.038                   & 0.000                 & 0.674                     & 0.037                   & 0.000                 & 0.692                    & 0.612                   & 0.741                   & 90\%                      & 0.738                    & 0.679                   & 0.789                   & 90\%                      & 0.832                    & 0.809                   & 0.854                   & 90\%                      \\
\multicolumn{1}{c}{}                           & Intercept                                                       & 0.122                     & 0.060                   & 0.043                 & 0.112                     & 0.060                   & 0.060                 & 0.090                     & 0.059                   & 0.131                 & 0.195                     & 0.060                   & 0.001                 & 0.123                     & 0.059                   & 0.038                 & 0.071                    & -0.063                  & 0.213                   & 21\%                      & -0.013                   & -0.165                  & 0.113                   & 2\%                       & -0.401                   & -0.666                  & -0.038                  & 89\%                      \\
\multicolumn{1}{c}{}                           & N                                                               &                           &                         & 6000                  &                           &                         & 6000                  &                           &                         & 6000                  &                           &                         &                       &                           &                         & 6000                  &                          &                         &                         &                           &                          &                         &                         &                           &                          &                         &                         &                           \\
\multicolumn{1}{c}{}                           & R2                                                              &                           &                         & 0.431                 &                           &                         & 0.432                 &                           &                         & 0.432                 &                           &                         & 0.436                 &                           &                         & 0.434                 &                          &                         &                         &                           &                          &                         &                         &                           &                          &                         &                         &                           \\
\midrule
\multirow{12}{*}{\rotatebox[origin=c]{90}{Success}}                       & Female                                                          & -0.124                    & 0.050                   & 0.013                 & -0.117                    & 0.051                   & 0.020                 & -0.149                    & 0.051                   & 0.003                 & -0.074                    & 0.051                   & 0.147                 & -0.133                    & 0.051                   & 0.009                 & -0.123                   & -0.243                  & -0.007                  & 61\%                      & -0.151                   & -0.262                  & -0.008                  & 74\%                      & -0.151                   & -0.297                  & 0.047                   & 64\%                      \\
                                               & Femaleness                                                      & -0.812                    & 0.067                   & 0.000                 & -0.803                    & 0.066                   & 0.000                 & -0.839                    & 0.067                   & 0.000                 & -0.944                    & 0.067                   & 0.000                 & -0.931                    & 0.068                   & 0.000                 & -0.831                   & -1.021                  & -0.533                  & 90\%                      & -0.774                   & -1.049                  & -0.556                  & 90\%                      & -0.223                   & -0.926                  & 0.315                   & 59\%                      \\
                                               & \begin{tabular}[c]{@{}r@{}}Female:\\ Femaleness\end{tabular}    & 0.311                     & 0.094                   & 0.001                 & 0.297                     & 0.093                   & 0.002                 & 0.359                     & 0.094                   & 0.000                 & 0.278                     & 0.095                   & 0.003                 & 0.370                     & 0.095                   & 0.000                 & 0.274                    & 0.013                   & 0.455                   & 80\%                      & 0.260                    & 0.050                   & 0.427                   & 61\%                      & 0.071                    & -0.345                  & 0.345                   & 14\%                      \\
                                               & Tenure                                                          & 0.139                     & 0.004                   & 0.000                 & 0.139                     & 0.004                   & 0.000                 & 0.139                     & 0.004                   & 0.000                 & 0.138                     & 0.004                   & 0.000                 & 0.139                     & 0.004                   & 0.000                 & 1.191                    & 1.172                   & 1.216                   & 90\%                      & 1.180                    & 1.156                   & 1.205                   & 90\%                      & 1.155                    & 1.145                   & 1.178                   & 90\%                      \\
                                               & \begin{tabular}[c]{@{}r@{}}N. projects\\ (log)\end{tabular}     & 1.192                     & 0.022                   & 0.000                 & 1.192                     & 0.022                   & 0.000                 & 1.192                     & 0.022                   & 0.000                 & 1.207                     & 0.022                   & 0.000                 & 1.189                     & 0.022                   & 0.000                 & 0.139                    & 0.137                   & 0.141                   & 90\%                      & 0.140                    & 0.137                   & 0.142                   & 90\%                      & 0.144                    & 0.141                   & 0.144                   & 90\%                      \\
                                               & \begin{tabular}[c]{@{}r@{}}Total activity\\ (log)\end{tabular}  & 0.505                     & 0.042                   & 0.000                 & 0.511                     & 0.042                   & 0.000                 & 0.525                     & 0.042                   & 0.000                 & 0.475                     & 0.042                   & 0.000                 & 0.516                     & 0.042                   & 0.000                 & 0.531                    & 0.469                   & 0.570                   & 90\%                      & 0.562                    & 0.518                   & 0.598                   & 90\%                      & 0.634                    & 0.607                   & 0.661                   & 90\%                      \\
                                               & Intercept                                                       & -0.035                    & 0.067                   & 0.605                 & -0.045                    & 0.067                   & 0.496                 & -0.054                    & 0.066                   & 0.416                 & 0.058                     & 0.067                   & 0.386                 & -0.002                    & 0.066                   & 0.979                 & -0.054                   & -0.212                  & 0.074                   & 8\%                       & -0.105                   & -0.244                  & 0.034                   & 32\%                      & -0.425                   & -0.721                  & -0.052                  & 88\%                      \\
                                               & N                                                               &                           &                         & 6000                  &                           &                         & 6000                  &                           &                         & 6000                  &                           &                         & 6000                  &                           &                         & 6000                  &                          &                         &                         &                           &                          &                         &                         &                           &                          &                         &                         &                           \\
                                               & R2                                                              &                           &                         & 0.474                 &                           &                         & 0.474                 &                           &                         & 0.474                 &                           &                         & 0.478                 &                           &                         & 0.476                 &                          &                         &                         &                           &                          &                         &                         &                           &                          &                         &                         &                           \\
                                               \midrule
\multirow{13}{*}{\rotatebox[origin=c]{90}{Survival}}                
                                               & Female                                                          & 0.042                     & 0.142                   & 0.768                 & 0.045                     & 0.144                   & 0.752                 & -0.012                    & 0.144                   & 0.932                 & 0.005                     & 0.147                   & 0.975                 & -0.020                    & 0.144                   & 0.892                 & -0.307                   & -0.515                  & -0.052                  & 52\%                      & -0.307                   & -0.626                  & -0.040                  & 45\%                      & -0.243                   & -0.676                  & 0.276                   & 25\%                      \\
                                               & Femaleness                                                      & -0.637                    & 0.197                   & 0.001                 & -0.718                    & 0.192                   & 0.000                 & -0.769                    & 0.196                   & 0.000                 & -0.640                    & 0.196                   & 0.001                 & -0.690                    & 0.196                   & 0.000                 & -0.727                   & -1.051                  & -0.403                  & 90\%                      & -0.654                   & -1.068                  & -0.259                  & 81\%                      & -0.218                   & -0.787                  & 0.449                   & 18\%                      \\
                                               & \begin{tabular}[c]{@{}r@{}}Female:\\ Femaleness\end{tabular}    & 0.422                     & 0.269                   & 0.118                 & 0.445                     & 0.268                   & 0.097                 & 0.557                     & 0.270                   & 0.039                 & 0.478                     & 0.273                   & 0.080                 & 0.539                     & 0.271                   & 0.047                 & 0.489                    & 0.057                   & 0.854                   & 33\%                      & 0.436                    & -0.089                  & 1.025                   & 23\%                      & 0.175                    & -0.849                  & 1.043                   & 9\%                       \\
                                               & Tenure                                                          & -0.105                    & 0.011                   & 0.000                 & -0.105                    & 0.011                   & 0.000                 & -0.105                    & 0.011                   & 0.000                 & -0.104                    & 0.011                   & 0.000                 & -0.104                    & 0.011                   & 0.000                 & 1.165                    & 1.148                   & 1.194                   & 90\%                      & 1.156                    & 1.140                   & 1.188                   & 90\%                      & 1.138                    & 1.129                   & 1.154                   & 90\%                      \\
                                               & \begin{tabular}[c]{@{}r@{}}N. projects\\ (log)\end{tabular}     & 1.129                     & 0.072                   & 0.000                 & 1.133                     & 0.072                   & 0.000                 & 1.130                     & 0.072                   & 0.000                 & 1.128                     & 0.073                   & 0.000                 & 1.123                     & 0.072                   & 0.000                 & -0.110                   & -0.112                  & -0.108                  & 90\%                      & -0.109                   & -0.112                  & -0.107                  & 90\%                      & -0.106                   & -0.108                  & -0.106                  & 90\%                      \\
                                               & \begin{tabular}[c]{@{}r@{}}Total activity\\ (log)\end{tabular}  & 1.283                     & 0.131                   & 0.000                 & 1.271                     & 0.131                   & 0.000                 & 1.285                     & 0.130                   & 0.000                 & 1.288                     & 0.131                   & 0.000                 & 1.299                     & 0.130                   & 0.000                 & 1.430                    & 1.381                   & 1.464                   & 90\%                      & 1.453                    & 1.370                   & 1.492                   & 90\%                      & 1.503                    & 1.472                   & 1.526                   & 90\%                      \\
                                               & Intercept                                                       & -3.406                    & 0.201                   & 0.000                 & -3.355                    & 0.200                   & 0.000                 & -3.356                    & 0.199                   & 0.000                 & -3.410                    & 0.202                   & 0.000                 & -3.405                    & 0.199                   & 0.000                 & -3.310                   & -3.469                  & -3.127                  & 90\%                      & -3.358                   & -3.563                  & -3.086                  & 90\%                      & -3.597                   & -3.935                  & -3.314                  & 90\%                      \\
                                               & N                                                               &                           &                         & 6000                  &                           &                         & 6000                  &                           &                         & 6000                  &                           &                         & 6000                  &                           &                         & 6000                  &                          &                         &                         &                           &                          &                         &                         &                           &                          &                         &                         &                           \\
                                               & AIC                                                             &                           &                         & 14583                 &                           &                         & 14579                 &                           &                         & 14578                 &                           &                         & 14584                 &                           &                         & 14582                 &                          &                         &                         &                           &                          &                         &                         &                           &                          &                         &                         &                           \\
                                               & BIC                                                             &                           &                         & 14635                 &                           &                         & 14631                 &                           &                         & 14630                 &                           &                         & 14635                 &                           &                         & 14634                 &                          &                         &                         &                           &                          &                         &                         &                           &                          &                         &                         &  \\
                           \bottomrule
\end{tabular}
\end{adjustbox}
\caption{Behance models. OLS Regression Model tables predicting Log(Attention), Log(Success), and logistic regression models predicting staying active on Behance one year after data collection. All variables are normalized between 0 and 1. Samples indicate independent samples containing 6,000 female and male Behance users. Robustness results are run on 100 5\%, 10\%, and 25\% randomly gender-swapped datasets. Avg. shows the average coefficients of the models, Min and Max indicate the lowest and highest values of 100 runs, and Sign. shows the ratio of significant coefficients.}
\label{tab:Model_behance}
\end{sidewaystable}\textbf{}

\begin{table}[]
\begin{tabular}{lrrrrrr}
\toprule
                                                                                                         & \multicolumn{3}{c}{GitHub}                                                                 & \multicolumn{3}{c}{Behance}                                                                \\
                                                                                                    & \multicolumn{1}{c}{Attention} & \multicolumn{1}{c}{Success} & \multicolumn{1}{c}{Survival} & \multicolumn{1}{c}{Attention} & \multicolumn{1}{c}{Success} & \multicolumn{1}{c}{Survival} \\
\midrule
\begin{tabular}[c]{@{}l@{}}Males' prediction at male's\\ femaleness median\end{tabular}                  & 39.24                         & 1.67                        & 0.94                         & 180.69                        & 313.44                      & 0.44                         \\
\begin{tabular}[c]{@{}l@{}}Males' prediction at female's\\ femaleness median\end{tabular}                & 22.70                         & 1.32                        & 0.92                         & 87.02                         & 171.09                      & 0.39                         \\
\begin{tabular}[c]{@{}l@{}}Females' prediction at male's\\ femaleness median\end{tabular}                & 52.16                         & 1.63                        & 0.92                         & 210.91                        & 333.61                      & 0.42                         \\

\begin{tabular}[c]{@{}l@{}}Females' prediction at\\ female's femaleness median\end{tabular}              & 37.62                         & 1.24                        & 0.88                         & 113.81                        & 221.01                      & 0.39                         \\
\midrule
Total female disadvantadge                                                                               & 1.62                          & 0.43                        & 0.06                         & 66.88                         & 92.43                       & 0.05                         \\
Female's indirect discrimination                                                                         & 14.54                         & 0.39                        & 0.04                         & 97.10                         & 112.60                      & 0.03                         \\
Male's indirect discrimination                                                                           & 16.54                         & 0.35                        & 0.02                         & 93.67                         & 142.35                      & 0.05                         \\
Direct discrimination                                                                                    & -12.92                        & 0.04                        & 0.01                         & -30.22                        & -20.17                      & 0.02                         \\
\begin{tabular}[c]{@{}l@{}}\% indirect discrimination out of total \\ disadvantage of women\end{tabular} & 898\%                      & 90\%                     & 74\%                      & 145\%                      & 122\%                    & 60\%                      \\
\begin{tabular}[c]{@{}l@{}}\% direct discrimination out of total \\ disadvantage of women\end{tabular}   & -798\%                     & 10\%                     & 26\%                      & -45\%                      & -22\%                    & 40\%                     \\
\bottomrule
\end{tabular}
\caption{Medians of Attention, Success and Survival by males' and female's femaleness median, and quantified direct and indirect discrimination on GitHub and Behance. Total female disadvantage is calculated as the difference between predicted outcome values of men at males' femaleness median, and women's at females' femaleness median. Indirect gender discrimination is the slope of women's predicted outcome along the values of femaleness, specifically the difference between the the predicted outcome value of women at  males' and female's femaleness median. Direct discrimination is the difference between the predicted outcome values of men and women at male's femaleness median.}
\label{tab:disadvantage}
\end{table}

\end{document}